\documentclass[]{jpconf}
\usepackage{graphicx}
\begin{document}
\title{Non-Boltzmann Ensembles and Monte Carlo Simulations}
\author{K. P. N. Murthy}
\address{School of Physics, University of Hyderabad, Hyderabad 500 046 Telengana India\\[2mm]
and\\[2mm]
Manipal Centre for Natural Sciences,
Manipal University,
Manipal  576 104, Karnataka, India}
\ead{kpnmsp@uohyd.ac.in}
\begin{abstract}Boltzmann sampling based on Metropolis algorithm 
has been extensively used for simulating a canonical ensemble and estimating 
macroscopic properties of a closed system at the desired temperature. 
An estimate of a mechanical property, like energy, of an 
equilibrium system, is made by averaging over  
a large number microstates generated by Boltzmann  Monte Carlo methods.
This is possible because we can assign a numerical value for energy to each microstate.  
However, a thermal property like entropy, is not easily accessible to 
these methods. The reason is simple. We can not assign a numerical 
value for entropy, to a  microstate. Entropy is not a property associated 
with any single  microstate.
It is a collective property of all
the microstates.  Toward calculating entropy and other thermal 
properties,  a non-Boltzmann Monte Carlo technique called  Umbrella 
sampling was proposed some forty years ago. Umbrella sampling 
has since undergone several metamorphoses and we have now,  multi-canonical 
Monte Carlo, entropic sampling, flat histogram methods, 
Wang-Landau algorithm {\it etc.} This class of methods generates 
non-Boltzmann ensembles which are un-physical. However, physical 
quantities can be calculated as follows. 
First  un-weight  a microstates of the entropic ensemble; then re-weight it    
to the desired physical ensemble. Carry out weighted average over the entropic ensemble
to estimate physical quantities.
In this talk I shall tell
you of the most recent  non-Boltzmann Monte Carlo method and 
show how to calculate free energy  for a few systems.
We first consider  estimation of  free energy as a function of energy at different temperatures
to characterize phase transition in an hairpin DNA 
in the presence of an unzipping force. 
Next we consider free energy as a function of order parameter and  
to this end we estimate
density of states $g(E,M)$, as   a function of both energy $E$,  and order parameter $M$.
This is carried out in two stages. We estimate $g(E)$ in the first stage. 
Employing $g(E)$, we generate an entropic ensemble. In the second stage,  
we estimate $g(E,M)$ by averaging 
$g(E)$ over partial  entropic ensemble for various values of the order parameter $M$.  
We present results on 
Ising spin system which exhibits second order  phase transition and on 
a double strand DNA, which exhibits  first order phase  transition.  
\end{abstract}
\section{Introduction}
Corresponding to a thermodynamic property, we have, in statistical mechanics, a random variable.
The average of the random variable over a suitable statistical ensemble, {\it e.g.} microcanonical  for an 
isolated system, canonical for a closed system,  and  grand canonical for an open system,  
gives the value of the corresponding thermodynamic property. For example, consider internal energy 
$U$. Corresponding to this, in statistical mechanics, we have energy
$E$ - the kinetic energy and interaction energy of the atoms and molecules
of the macroscopic object. A numerical value of $E$ can be assigned to each microstate. 
The value of $E$ fluctuates when an equilibrium system switches 
from one microstate to another. These fluctuations are  an integral part of an  
equilibrium description. The average of $E$ over a canonical 
ensemble at a given temperature  equals $U$.
The computation of average energy is then straight forward.
Generate a canonical  ensemble  employing Monte Carlo method.  
A simple arithmetic average of energy over a Monte Carlo sample  gives the 
required answer. The statistical error associated with the estimated average is also computed from the same sample. 
Such a neat computational scheme is possible because a numerical value for energy can be 
assigned to each microstate.  

How does one calculate entropy ? We notice  that we can not assign a  value of entropy to any 
single microstate. Entropy is a property that 
belongs collectively to all the  microstates. While energy is a \lq\lq\  private \rq\rq\   property, 
entropy is a \lq\lq\ social \rq\rq\ 
 or \lq\lq\ public \rq\rq\  property, see below.
 
Let $\{ C_i\ :\ i=1,\  2,\  \cdots\}$ denote the microstates of an equilibrium system and 
$\{ p(C_i)\ :\ i=1,\ 2,\ \cdots\}$, the corresponding probabilities. The Boltzmann-Gibbs-Shannon entropy is given by 
$
S=-k_B\sum_i p(C_i)\ln p(C_i).$
For an isolated system, $p(C_i)=1/\widehat{\Omega}(E,V,N)$, 
where $\widehat{\Omega}$ denotes the number of microstates. For a simple fluid  $\widehat{\Omega}$ depends on
energy $E$, volume $V$, and number of particles $N$. 
We have  $S=k_B\ln\widehat{\Omega}(E,V,N)$. 
For a closed system at temperature 
$T=1/[k_B\beta]$, we have, 
$
p(C_i)=Q^{-1}\exp[-\beta E(C_i)],
$
where $
Q(T, V, N)=\sum_i\exp[-\beta E(C_i)]
$
  is the canonical partition function. 

\section{Metropolis Algorithm}
Consider a  system
characterized by probabilities $\{ p(C_\nu)\ :\ \nu=1,2,\cdots\}$. Our aim is to generate a large number of  
microstates  of the system consistent with the given probabilities.
 To this end, start with an arbitrary initial microstate $C_0$ and  generate a Markov chain,
$C_0 \to C_1 \to \cdots C_i \to C_{i+1} \cdots\ ,
$  
employing  Metropolis rejection algorithm\cite{MRRTT}, see below.
 
Let $C_i$ be the current microstate and $p_i=p(C_i)$, its 
probability; we make a change in the current microstate and construct a 
trial microstate $C_t$. For example, if we are simulating 
Ising spin system, select a spin randomly from the current spin configuration,  
and flip it to get a  trial microstate.  
Let $p_t=p(C_t)$. Calculate 
$
p=\ {\rm min.} ( 1,p_t/p_i).$
Generate a random number $\xi$ uniformly and independently
distributed between zero and unity. If $\xi\ \le\ p$ 
accept the trial state and advance the Markov chain to $C_{i+1}=C_t$. If not,  
reject the trial state and advance the Markov chain to  $C_{i+1}=C_i$. 
Repeat the process on the microstate $C_{i+1}$ and 
proceed to generate a long Markov chain. 
The asymptotic part of the chain 
shall contain microstates 
belonging to an ensemble 
characterized by  $\{ p(C_\nu)\ :\ \nu=1,2,\cdots\}$.
\section{Boltzmann Sampling}
When the probabilities $\{ p(C_\nu)\ :\ \nu=1,2.\cdots\}$ describe a physical ensemble {\it e.g.} canonical ensemble or grand canonical ensemble,
we call it Boltzmann sampling. 
Notice  Metropolis algorithm  requires only 
the ratio of probabilities;
hence   we need to know $p(C)$ only upto a normalization constant.
It is precisely because of this we can simulate a closed system, since we 
need to know only the Boltzmann weight, $\exp[-\beta E(C)]$, 
and  not the canonical partition
function. Also the algorithm obeys detailed balance 
and hence generates a reversible Markov
chain, see {\it e.g.} \cite{KPNVSS}, appropriate for describing an equilibrium system.
 
Generate a Markov chain until it equilibrates;
continue the Markov chain and  collect a set of large number of microstates 
$\{ C_i\ : i=1,2,\cdots M\}$.
Let $O$ be a property of interest and  $O(C)$,  its value,  when the system 
in a microstate $C$. Then,
\begin{eqnarray}
\langle O\rangle &=&{}^{\ {\rm Limit}}_{M\to\infty}\ \frac{1}{M}\sum_{i=1}^M O(C_i). 
\end{eqnarray} 
To calculate the statistical error associated with our estimate of 
the mean, we proceed as follows.
We calculate  the second moment 
\begin{eqnarray}
\langle O^2\rangle  &=&{}^{\ Limit}_{M\to\infty}\ 
 \frac{1}{M}\sum_{i=1}^M O^2(C_i),
\end{eqnarray}
and  the variance $\sigma^2=\langle O^2\rangle -\langle O\rangle ^2$.
The statistical error is then   given by $\sigma/\sqrt{M}$.
Notice that  the Boltzmann sampling described above, depends 
crucially on our ability to assign a numerical value
of the property  $O$
to each microstate of the system.

Consider estimating  a property like entropy. We can not assign a numerical
value for entropy to a microstate. Entropy is a collective property of all the microstates
of an  equilibrium system. Hence entropy can not be easily obtained from Boltzmann sampling. 
We need non-Boltzmann sampling techniques and to this, we turn our attention below. 
\section{Non-Boltzmann Sampling}
Torrie and Valleau\cite{TV} were, perhaps, the first to propose a non-Boltzmann algorithm to calculate thermal properties. 
Their method, called Umbrella sampling,  has since undergone a series of metamorphoses. 
We have multi-canonical Monte Carlo algorithm of Berg and Neuhaus\cite{BN}, entropic sampling of Lee\cite{L}
and the algorithm of Wang and Landau\cite{WL}. We describe below the algorithm due to Wang and Landau\cite{WL}.

Let $\widehat{\Omega}(E)$  denote the density of states of the system under simulation.
The microcanonical entropy is given by logarithm of the density of states : $S(E)=\ln \widehat{\Omega}(E)$. 
Let $C$ be a microstate  and $E_C=E(C)$,  its energy. We define 
an ensemble characterized by the probabilities $p(C)\propto 1/\widehat{\Omega}(E_C),$ defined for 
all the microstates of the system. We employ Metropolis rejection technique to generate microstates 
based on these probabilities. The probability of acceptance of a trial microstate is  given by,
$p={\rm min.} (1,p_t/p_i) ={\rm min.} [ 1,\widehat{\Omega}(E_i)/\widehat{\Omega}(E_t)].$
Note, if the trial state belongs to a low entropy region, it gets accepted with unit probability; if not, 
its acceptance probability is less than unity. Thus, in the rejection step, low entropy regions   
are preferred, statistically.  This preference cancels statistically exactly, the natural tendency of  
random sampling trial states from  regions of high entropy. 
As a result,  the ensemble generated by the algorithm will 
have equal number of microstates in equal regions of energy.
In other words, the histogram of energy of 
the microstates of the ensemble shall be flat. 
But a crucial point remains :  we do not know 
the density of states $\widehat{\Omega}$, as yet.

Wang and Landau\cite{WL} proposed to estimate  
the density of states in an initial  learning run. 
We define  a function  $g(E)$ and set it to unity for all $E$. 
We also define a histogram of energy
$H(E)$ and set it to zero for all $E$. 
Start with an initial microstate $C_0$.
Let $E_0=E(C_0)$ be its energy. 
Update $g(E_0)$ to $\alpha\times g(E_0)$  where $\alpha$ is called 
the Wang-Landau factor, set to $\alpha_0=e$,  
in the first iteration. Also update  $H(E_0)$ 
to $H(E_0)+1$. Construct a  chain of microstates as per Metropolis rejection technique 
taking the probabilities proportional to $1/g(E(C))$. Every time you advance the  chain, 
update  $g$ and $H$. The chain will not be  Markovian  since  since the transition 
probabilities 
depend on the entire past. Carry out the simulation of the chain  
until the histogram is flat over a range of energy.  
Then set $\alpha=\alpha_1= \sqrt{\alpha_0}$, reset $H(E)=0\ \forall\  E$ and proceed with the second iteration of the 
learning run. The value of $\alpha$ tends to unity upon further iterations. 
After some twenty five iterations, $\alpha=\alpha_{25}\approx 1+3\times 10^{-7}$. The histogram would be 
flat over the range of energy of interest. Flatter the histogram, closer is $g(E)$ to 
the true density of states $\widehat{\Omega}(E)$.  We take  $g(E)$ at the end of last iteration of the learning 
run, that leads to a reasonably flat histogram,   as an estimate
of $\widehat{\Omega}(E)$. We can define suitable criteria for flatness of the histogram. For example we 
can say the histogram is flat if the smallest and largest entries in the histogram do not 
differ from each other by more than say 
$10\%$. Note we calculate the density of states only upto a multiplicative constant. 
The microcanonical entropy is given by 
$S(E)=\ln g(E)$, upto an additive constant. We can then employ the machinery of thermodynamics 
to calculate all the 
thermal properties, starting from the microcanonical entropy.
Also, the converged density of states can be  employed in the production run 
and a large number of microstates generated. These microstates belong to an ensemble which we call as 
entropic ensemble or Wang-Landau ensemble. The Wang-Landau  ensemble is un-physical. Nevertheless, physical 
quantities can be estimated by un-weighting and re-weighting, see below. 

We first attach a statistical weight of unity 
to each microstate of the Wang-Landau ensemble. 
Then we divide the statistical weight  by $p(C)=1/g(E(C))$. 
This is called un-weighting.
Upon un-weighting, the ensemble of weighted microstates  becomes microcanonical.
In other words, weighted averaging over the microstates of the (unphysical) 
Wang-Landau  ensemble is equivalent to averaging over the (physical) microcanonical ensemble. We can further   
re-weight  to the desired
ensemble. For example, to calculate the average over canonical ensemble, we multiply by the 
Boltzmann weight, see below. 
 
Let $\{ C_i : i=1,2,\cdots ,M\}$ denote the 
microstates of the Wang-Landau ensemble. 
 We carry out weighted average after un-weighting and re-weighting,
\begin{eqnarray}
\langle O\rangle &=& 
{}^{\ {\rm Limit}}_{M\to\infty}\ 
\frac{\sum_{i=1}^M O(C_i)W(C_i)}
           {\sum_{i=1}^M W(C_i)}
\end{eqnarray} 
where the weight factor is given by,
\begin{eqnarray}
W(C_i)&=&g(E(C_i))\exp(-\beta E(C_i))
\end{eqnarray}
$\langle O\rangle$  is the desired average over a canonical ensemble. 
                                                   
Thus, employing  Wang-Landau algorithm, mechanical as well as thermal properties can be calculated. 
So far so good. There are a few outstanding issues
with the Wang-Landau algorithm in particular and Non-Boltzmann Monte Carlo 
methods in general. The principal amongst them  
concerns the slow convergence of the 
density of states, for systems with continuous Hamiltonians, 
see {\it e.g.} \cite{JSM} for a discussion on this issue. 
Another issue is that the algorithm does not obey detailed balance.
There are also problems associated with calculation of statistical error bars.
We shall not address these and related problems in this talk. Instead
in what follows, we shall describe how does one calculate 
a thermal property like (Helmholtz) free energy
as a function of energy (or order parameter), for temperatures close 
to transition point, employing Wang-Landau algorithm. 
We shall also illustrate the technique on   a few examples.  
\section{Calculation of Free Energy}
The free energy of a closed system at temperature $T$ is given by,
$
F(T,V,N)=-k_B T \ln Q(T,V,N),
$
where $Q(T,V,N)$ is the canonical partition function. We can also define microcanonical free energy
for an isolated system as,
\begin{eqnarray}
F(U,V,N) = U(S,V,N)-{\displaystyle \left(\frac{\partial U}{\partial S}\right)_{V,N}\ S\ ,
}\end{eqnarray}
 where
$S(U,V,N)$ is the microcanonical entropy given by $S=k_B\ln \widehat{\Omega}(E,V,N)$. 

For an equilibrium system, free energy can not be simultaneously  
        a function of both energy and temperature :  an isolated system with a given energy has an unique 
        temperature; a closed system at a given temperature has a unique (average) energy. However we may be  interested
        in estimating the penalty in terms of excess free energy required to keep 
        a closed system in a non-equilibrium state with its energy $E$
        different from the equilibrium energy $U(T)$. To this end, following Joon Chang Lee\cite{JCL},
        we define a phenomenological free energy $F_L$, which is simultaneously
        a function of both energy and temperature. Let $F(T)$ denote the free energy 
        of the closed system at temperature $T$. We have $F_L(E,T)\ge F(T)$ for all energies, and  equality obtains when 
        $E=U(T)$.
        The excess free energy is given by $\Delta F= F_L (E,T)- F(T)$, where,
        \begin{eqnarray}
        {\displaystyle F_L(E,T) = -k_B T \ln  \sum_{C} g(E(C))\ \exp\left[ -\beta E(C)\right]\ \delta \left( E(C)-E\right)}
        .\end{eqnarray}
        The sum runs over all the microstates belonging to the entropic ensemble collected 
        during the production run of the Wang-Landau Monte Carlo simulation.
\subsection{Phase Transition in hpDNA : Free Energy versus Energy}        
        For purpose of illustration, we present the results on free energy profiles
     of a DNA-hairpin (hp-DNA) subjected to a constant unzipping force. DNA has two strands of equal length
     twisted around a common axis to form a double helix. Each strand can be considered
     as a chain of monomers or bases. When complementary bases at the two ends of a strand pair up,  
     hairpin structure occurs,  with a loop and stem. The structure can be closed  or 
     open. The transition from closed to open state is called denaturation
     or unbinding. If  denaturation occurs by virtue of temperature,
     it is called melting. 
    Force induced melting  has been extensively studied, see  \cite{SM} 
     and the references cited therein.
 We present results on chain of length $110$ units. We determine the 
     transition temperature from the peak of the specific heat profile. We take two temperatures 
     in the neighbourhood and on either side of the transition point, to calculate free energy profiles.
     We have considered four site occupation, bond fluctuation model\cite{CK}, on a two dimensional square 
     lattice. For details of the simulation see \cite{SM}. 
     
\begin{figure}[h]
\includegraphics[width=29pc]{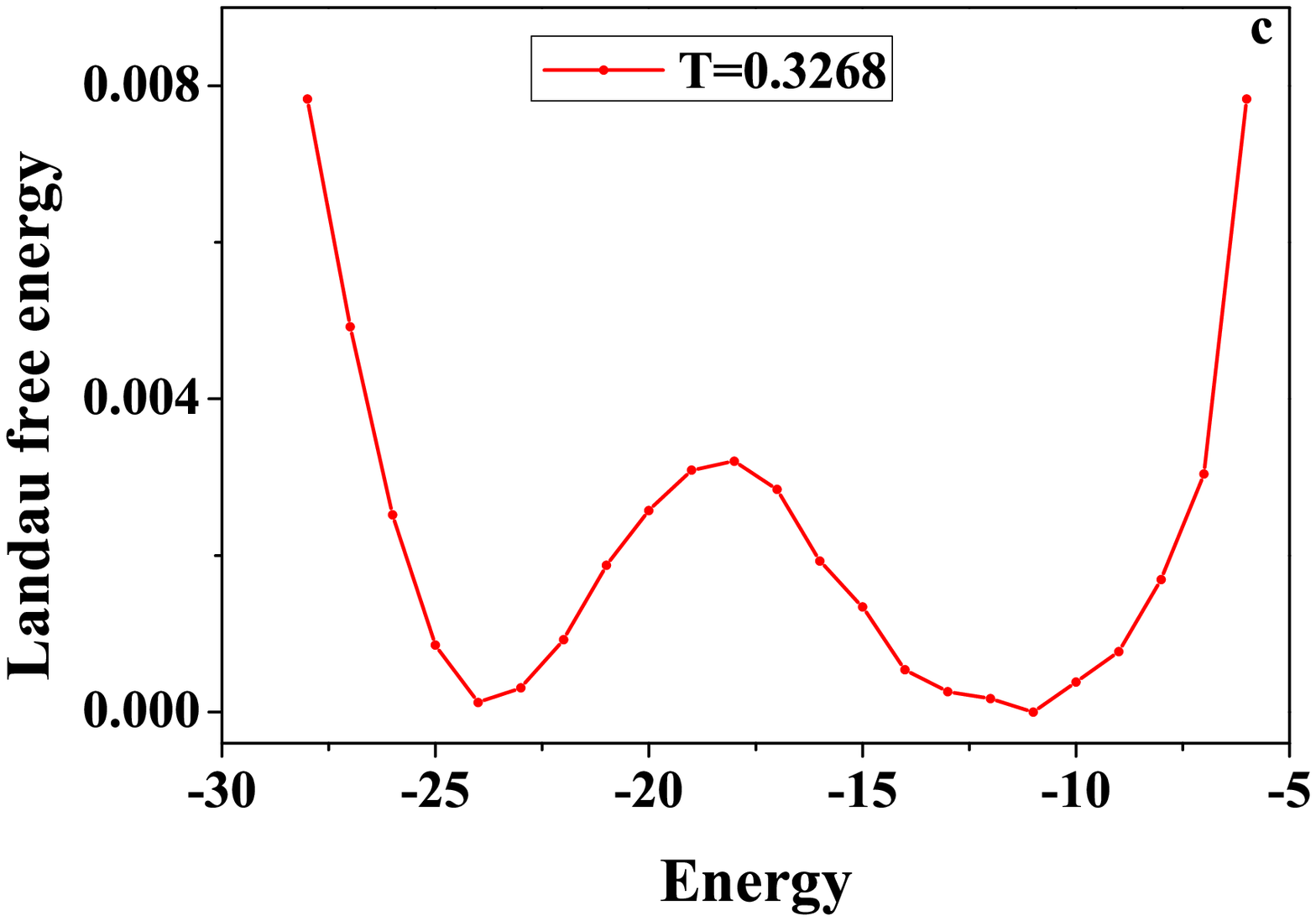}
\begin{minipage}[b]{08pc}
\caption{\label{label1} Variation of free energy with energy at  $T\ =\   T_C$ and unzipping force $f  = 0.04$.
Temperature at which the heat capacity peaks is taken as $T_C$. The chain length is $110$; reproduced 
from \cite{SM}}
\end{minipage}
\end{figure}

\begin{figure}[h]
\begin{minipage}{18pc}
\includegraphics[width=18pc]{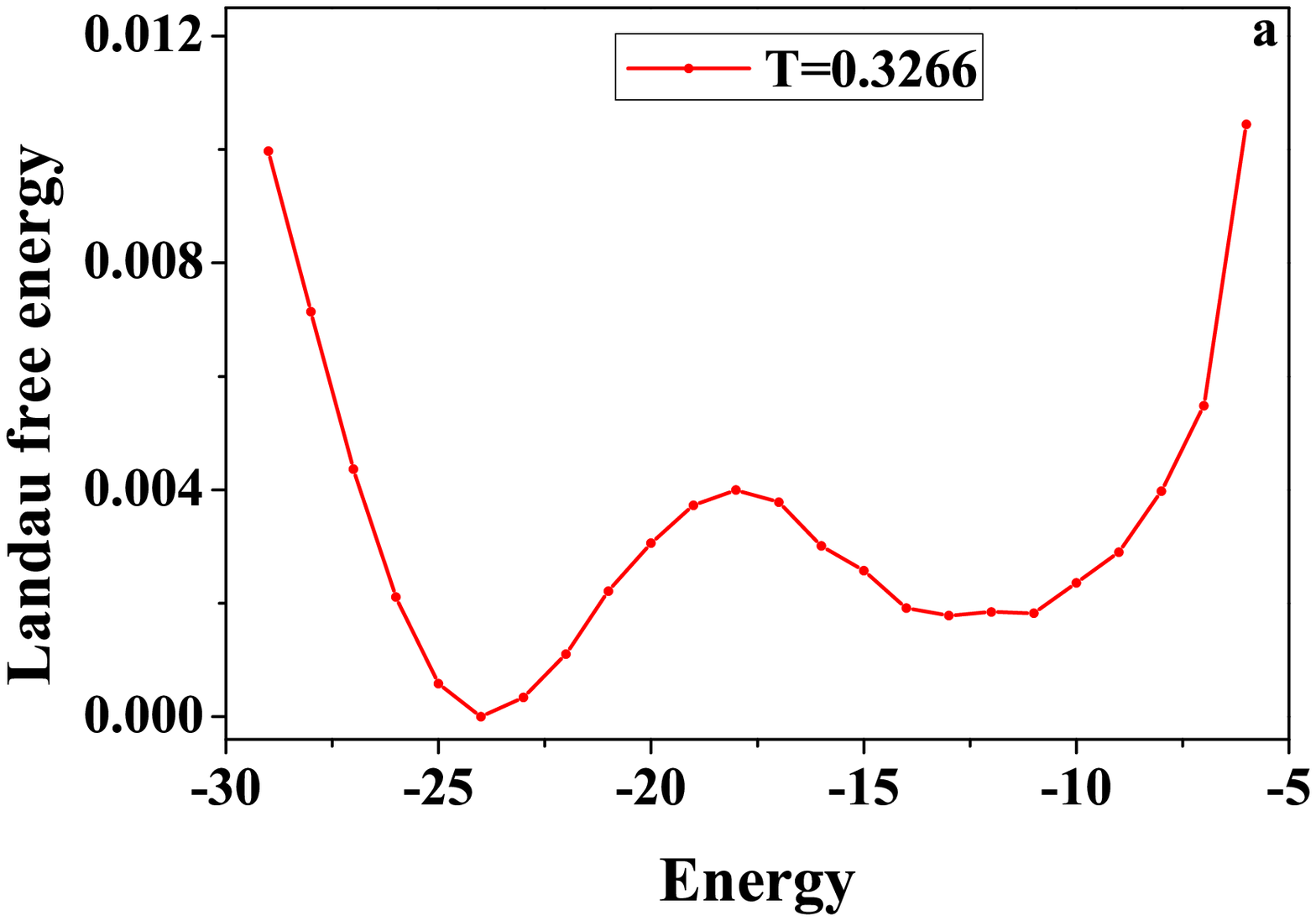}
\caption{\label{label2}Same as Figure 1, except\\ $T\ <\ T_C$; reproduced from \cite{SM}}
\end{minipage}\hspace{1pc}%
\begin{minipage}{18pc}
\includegraphics[width=18pc]{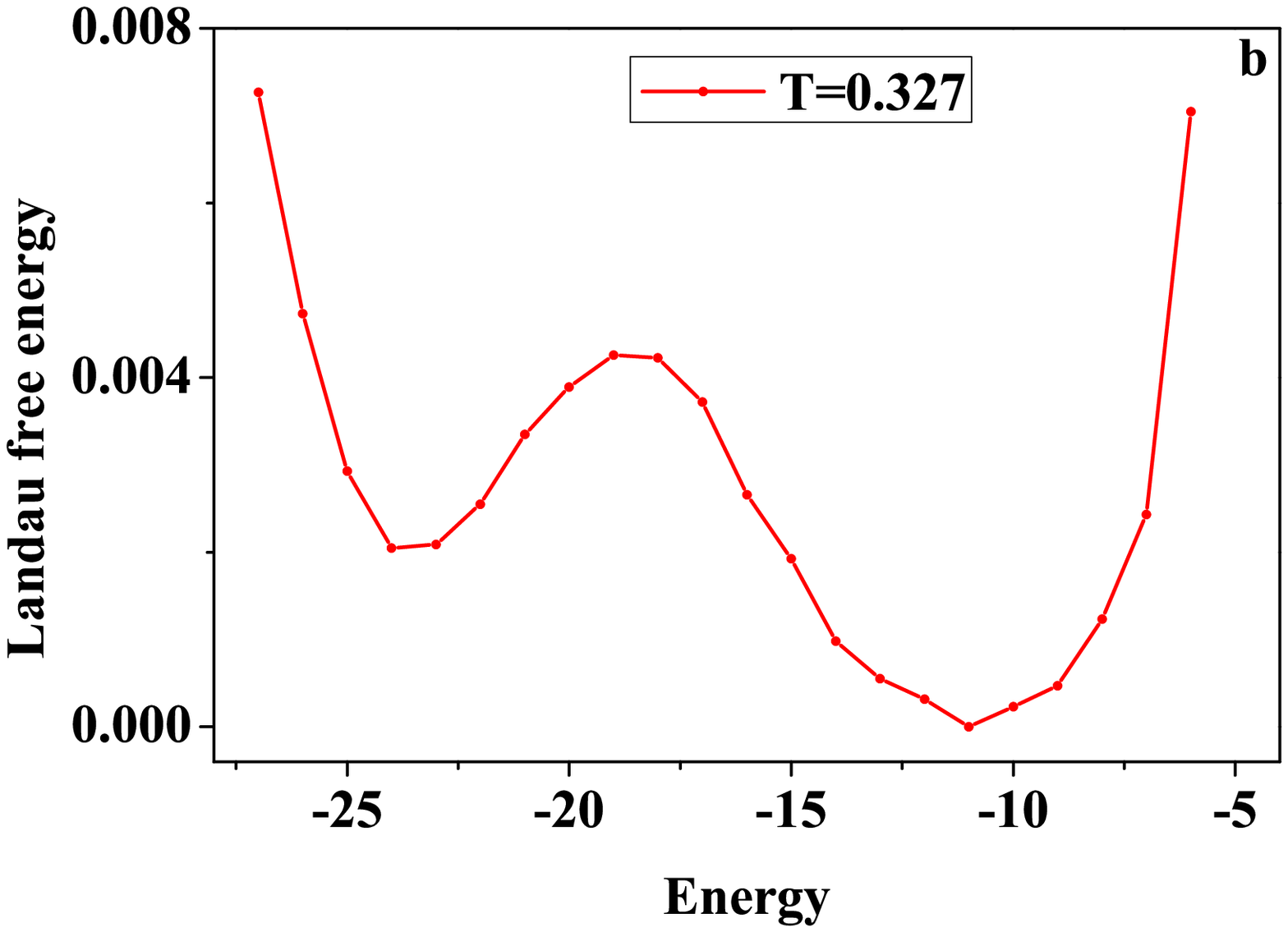}
\caption{\label{label3}Same as Figure 1, except\\  $T\ >\ T_C$; reproduced from \cite{SM}}
\end{minipage} 
\end{figure}

     We first calculate  heat capacity 
     and determine the temperature at which it is maximum. We take this as 
     transition temperature $T_C$. Then we estimate 
     the  free energy {\it versus} energy profiles and these are 
     depicted in Figures \ref{label1} - \ref{label3}, for $T=T_C$, $T\ <\ T_C$,  and $T\ >\ T_C$ respectively. 
     It is clear the transition is first order.
\section{Free Energy  versus Order Parameter}
Landau free energy is usually expressed as a function of order parameter $M$.   
Landau, Tsai, and Exler\cite{LTE} defined a joint density of states
$g(E,M)$ and employed it in the Wang-Landau algorithm. Let $E_i$ and $M_i$ be 
 the energy and order parameter of the current microstate; let
 $E_t$ and $M_t$ denote the same quantities  for the trial microstate. 
The acceptance probability 
in the Metropolis algorithm is taken as,
$
p={\rm min.} [1, g(E_i,M_i)/g(E_t,M_t)].
$
The joint density of states is updated at every step.
We also monitor the histogram $H(E,M)$. The histogram now  is a two dimensional surface.
When the surface becomes flat it indicates that $g(E,M)$ has converged to 
the true two dimensional density of states, $\widehat{\Omega}(E,M)$.      
     Obtaining a flat histogram in two dimensional space requires
considerable computing time\cite{LTE}. A good strategy\cite{GWLX}
 shall involve estimating  
the two dimensional density states in the production run, see below.
The learning run is reserved, as usual,  for obtaining a converged 
density of states in energy space only.

We assemble, in the production run,
an entropic ensemble. 
We first un-weight the microstate and then re-weight it to a 
microcanonical ensemble. We estimate  
the joint density of states as a weighted average over the entropic ensemble. This is  
 equivalent to averaging over a microcanonical ensemble. Thus, 
\begin{eqnarray}
\langle g(E_i,M_j) \rangle = 
\frac{ {\displaystyle \sum_C \delta(E(C)-E_i)\ \delta(M(C)-M_j)\ g(E(C))}}
{ {\displaystyle \sum_C \delta(E(C)-E_i)\ g(E(C))}} \hspace{12pt} \forall \ E_i,\ M_j.
\end{eqnarray}    
In the above, the sum runs over all the microstates of the entropic ensemble.
Having estimated  the joint density of states, 
we calculate the Landau free energy as
\begin{eqnarray}
F_L(M_j,T)=-k_BT \ln \sum_i \langle g(E_i,M_j)\rangle \exp(-\beta E_i)
\end{eqnarray}
where  we take $T$ to be very close to the transition temperature.

\subsection{Ising Spins, dsDNA and Order Parameter }
\begin{figure}[h]  
\includegraphics[width=28pc]{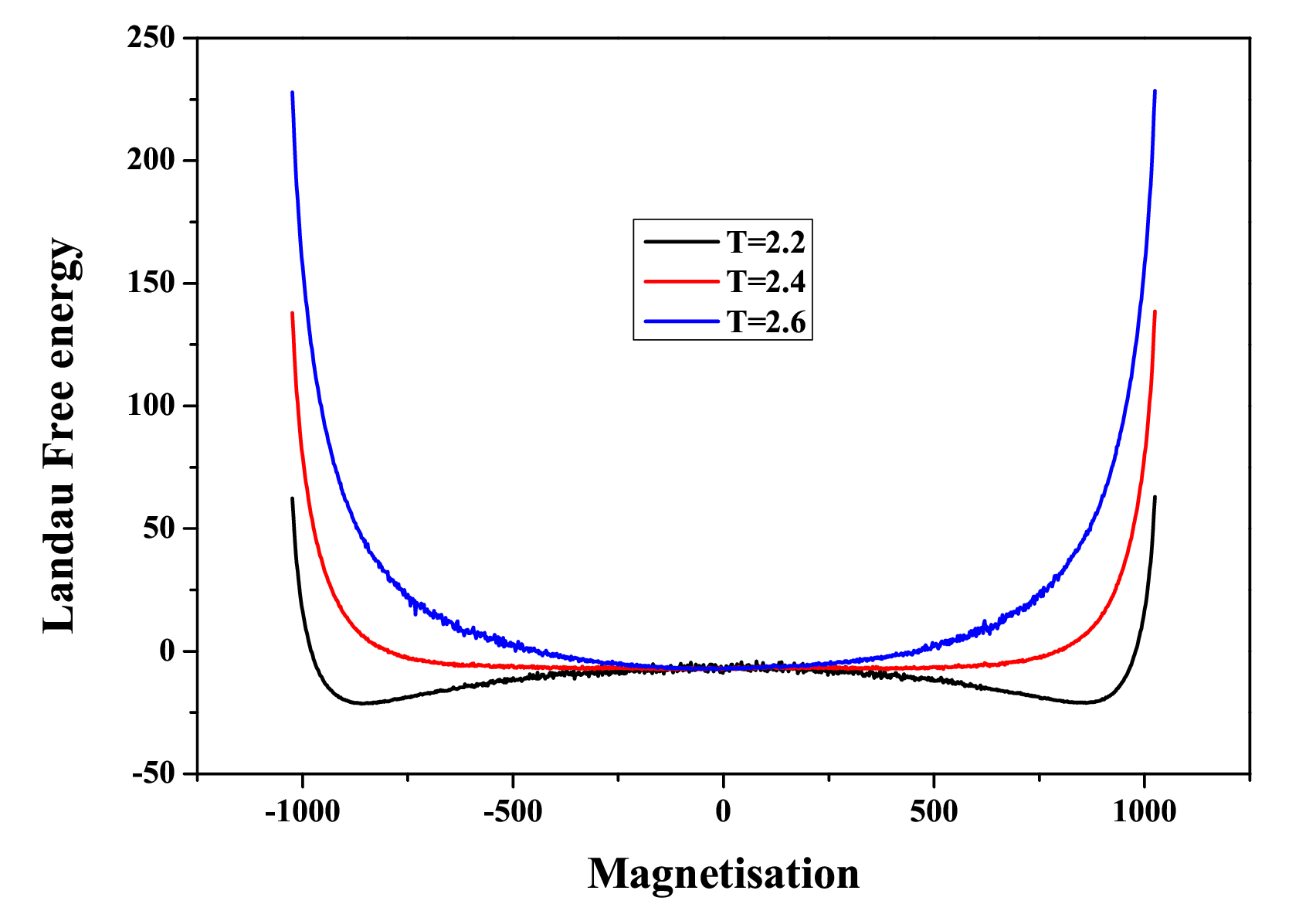}
\begin{minipage}[b]{09pc}\caption{\label{Ising32} Free energy {\it versus} magnetization for zero field 
Ising spin system; joint density of states has been calculated employing  the two stage method. 
Two dimensional square lattice with 
$L=32$. The top curve is for $T\ >\ T_C$; the middle curve is for $T=T_C$; and 
the bottom curve is for $T\ <\ T_C$;  reproduced from \cite{SBSM}}
\end{minipage}
\end{figure}

Figure \ref{Ising32} depicts the results on Landau  free energy, $F_L(M,T)$ versus magnetization, $M$, 
for a two dimensional Ising spin
 system. 
\noindent
The spins are located on the vertices of a $32\times 32$ 
square lattice. We have estimated the joint density of states, $g(E,M)$, 
in the production run of Wang-Landau simulation.  For details  see \cite{SBSM}.
Free energies  for 
$T\ <\ T_C\ ({\rm bottom\ curve}),\  T\ =\ T_C\ ({\rm middle\ curve}),\  {\rm and}\ T\ >\ T_C\ ({\rm top\ curve})$ 
are shown. 
The transition temperature $T_C$ was located from the specific heat profiles. 

\begin{figure}[h]
\includegraphics[width=27pc]{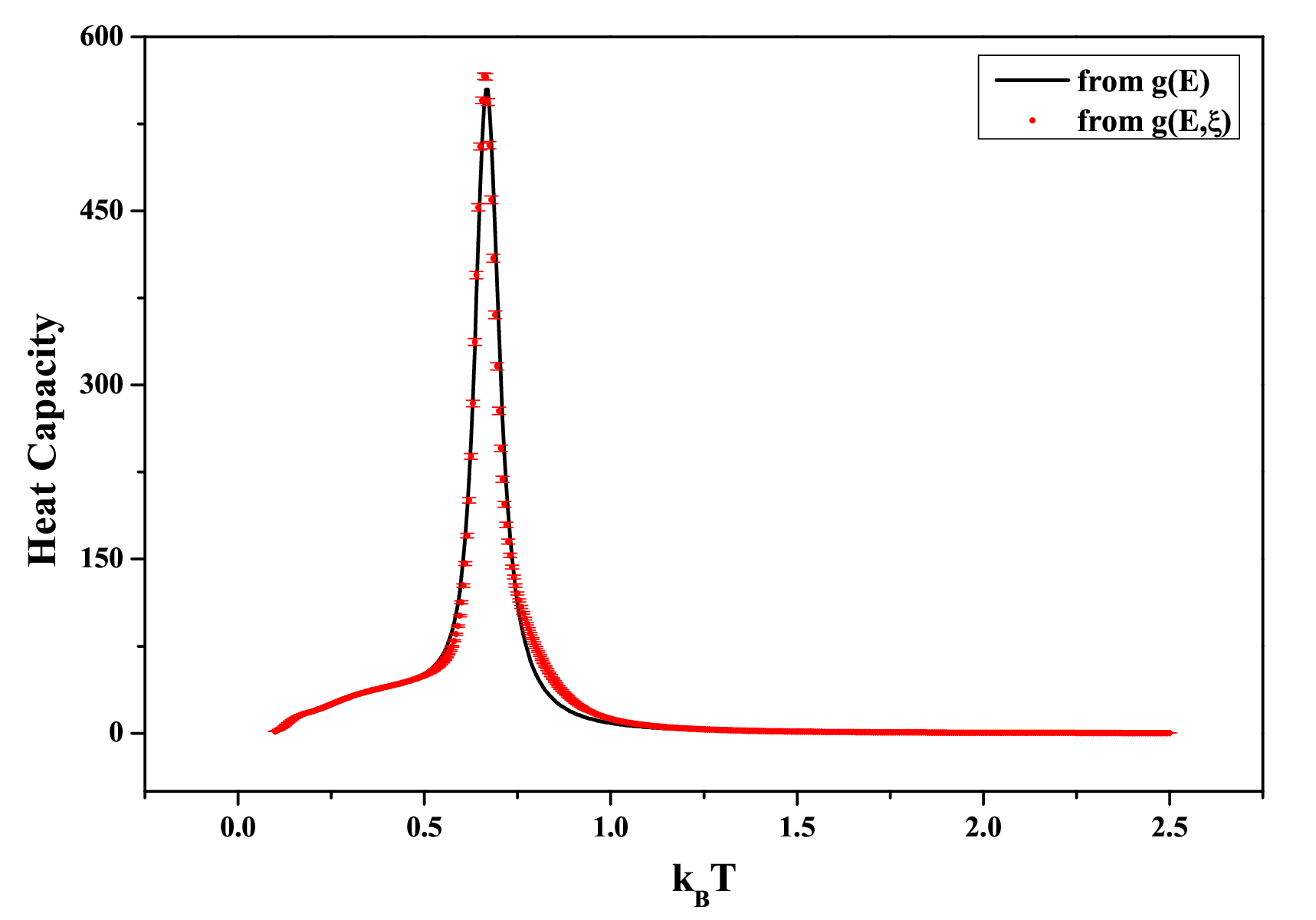}\hspace{1pc}%
\begin{minipage}[b]{10pc}
\caption{\label{cvdna} Heat capacity of dsDNA as a function of $T$ calculated from  
$g(E)$, estimated by  standard Wang-Landau algorithm and from $g(E,\xi)$, obtained  from the two stage method;
  reproduced from \cite{SBSM}}
\end{minipage}
\end{figure}

Figure \ref{cvdna}, reproduced from \cite{SBSM},  depicts results on variation 
of heat capacity with temperature, for double strand DNA. We have calculated 
free energy employing both the methods : (i) standard Wang-Landau algorithm in which $g(E,M)$ is obtained in the 
learning run and (ii) two-stage method in which $g(E,M)$ is estimated in the production run
by un-weighting and re-weighting procedures.  The two results agree
with each other, demonstrating that the two-stage method is as good 
as the standard Wang-Landau sampling  toward predicting mechanical properties.
I must mention that the two-stage method took considerably much less computer time.    

\begin{figure}[h]
\includegraphics[width=21pc]{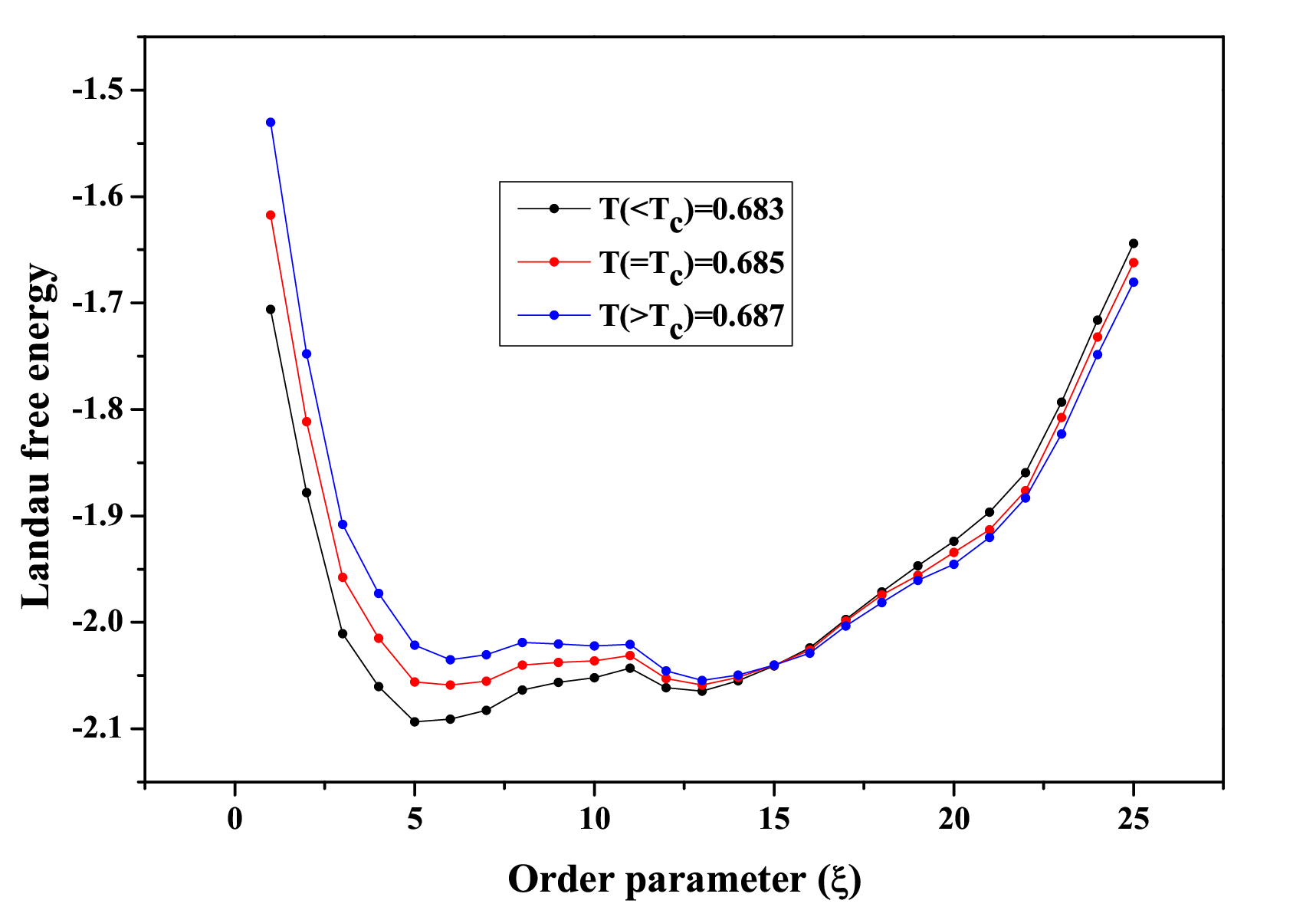}\hspace{1pc}%
\begin{minipage}[b]{16pc}
\caption{\label{lfedna} Free energy {\it versus} order parameter for a double strand DNA - dsDNA
of length $80$ in the presence of unzipping force $f=0.02$. The end to end distance is taken as 
order parameter. Joint density of states,  $g(E,\xi)$,  has been calculated 
employing the two stage method. Top curve is for $T\ <\ T_C$; 
the middle curve is for $T=T_C$; the bottom curve is for T$\ >\ T_C$;
 reproduced  from \cite{SBSM}}
\end{minipage} 
\end{figure}

Figure  \ref{lfedna} reproduced  from \cite{SBSM}, shows results on Landau free energy 
as a function of the order parameter for dsDNA. The end to end distance 
is taken as order parameter and is denoted by the symbol $\xi$. 
We have presented results 
at  $T=T_C$, $T\ <\ T_C$  and at $T\ >\ T_C$.
The transition is first order.

\noindent
\section{Summary}
I have introduced briefly the basic principles of non-Boltzmann Monte Carlo methods. 
The density of states is estimated through an iterative process. Logarithm of density of states 
gives entropy; all other thermal properties can be estimated from entropy, 
 employing the machinery of thermodynamics. Though a non-Boltzmann ensemble is unphysical,
we can extract averages of mechanical properties, over desired physical ensembles by employing un-weighting
and re-weighting techniques. Thus both thermal and mechanical properties
can be estimated employing non-Boltzmann Monte Carlo simulation methods. 
I have illustrated the usefulness of non-Boltzmann ensemble on three systems : an hairpin DNA,  Ising spins,
and double stranded DNA. Results on 
free energy profiles at temperatures close to transition point were presented.   
I believe it is time we take a Non-Boltzmann Monte Carlo,  as a method of first 
choice for simulating macroscopic systems.
\section*{Acknowledgements}
This talk is based on Ref. \cite{SM} and \cite{SBSM} which report work  carried out in collaboration with 
V S S Sastry, M Suman Kalyan, and  R Bharath; and to them I owe 
my thanks. Computations were carried out at CMSD, 
University of Hyderabad, Hyderabad,
and the Shakti cluster, MCNS, Manipal University, Manipal. 

\end{document}